# Design of Variable Bandpass Filters Using First Order Allpass Transformation And Coefficient Decimation

*S. J. Darak and A. P. Vinod*
School of Computer Engineering
Nanyang Technological University, Singapore
{dara0003, asvinod}@ntu.edu.sg

*E. M-K. Lai*
School of Engineering and Advanced Technology
Massey University, Albany, New Zealand
E.Lai@massey.ac.nz

***Abstract*— In this paper, the design of a computationally efficient variable bandpass digital filter is presented. The center frequency and bandwidth of this filter can be changed online without updating the filter coefficients. The warped filters, obtained by replacing each unit delay of a digital filter with an allpass filter, are widely used for various audio processing applications. However, warped filters fail to provide variable bandwidth bandpass responses for a given center frequency using first order allpass transformation. To overcome this drawback, our design is accomplished by combining warped filter with the coefficient decimation technique. The design example shows that the proposed variable digital filter is simple to design and offers a total gate count reduction of 36% and 65% over the warped filters compared to the designs presented in [3] and [1] respectively.**

## I. Introduction

In many audio signal processing applications, it is desirable to change the frequency response of a digital filter online with minimal overhead. Filters that allow one to do so are known as variable digital filters (VDFs). One particular type of VDF is the warped filter [1], obtained by replacing each unit delay of a digital filter with an allpass structure of an appropriate order. By changing the coefficients of the allpass structure, the magnitude response of the warped filter can be controlled on-the-fly. Warped filters are widely used for various audio applications such as linear prediction, echo cancellation, loudspeaker equalization, spectrally modifying an audio signal and detection of bandpass signals in a broadband signals [2-4]. Depending on the application, warped filters with variable lowpass, highpass, bandpass or bandstop responses can be designed. In this paper, the discussion is focused on warped filters with variable bandpass responses.

The ideal approach to design warped filter with variable bandpass response is to replace each unit delay with a second order allpass transformation. This approach is used in [3] to design a filter bank for low power digital hearing aids. It consists of variable lowpass, bandpass and highpass filters in parallel. However, the complexity of the bandpass filter, which uses second order allpass transformation, almost double than that of filters where first order allpass transformation is used. In [4], adaptive filters are designed using warping technique to detect the bandpass signals in a broadband signals. The adaptive filters are warped filters designed using reduced second order transformation and hence they provide fixed bandwidth bandpass responses at an arbitrary center frequency. Whenever the bandwidth needs to be changed, the filter coefficients will need to be updated, which incurs a large number of memory read and write operations.

In [5], a technique called coefficient decimation (CDM) is proposed for realizing low complexity reconfigurable VDFs with fixed coefficients. The CDM technique is combined with interpolation and masking techniques to realize reconfigurable filters in [6]. However, these methods are suitable only for designing discretely tunable filters, i.e. the set of distinct cut-off frequencies that can be achieved is finite.

In this paper, we present a low complexity VDF by combining warped filter with the CDM technique. The proposed VDF provides variable bandwidth bandpass responses for an arbitrary center frequency. The design example shows that the proposed VDF offers a substantial reduction in gate counts over other methods. The complexity of the filter bank in [3] can be reduced significantly when proposed VDF is used. Also, the proposed VDF can be used to extend the application in [4] for detection of variable bandwidth bandpass responses on-the-fly without the need of updating filter coefficients.

The rest of this paper is organized as follows. A review of W-FIR filter and CDM technique are provided in Section II. In Section III, our proposed design method for the VDF is presented. A design example and its implementation results are discussed in Section IV and V respectively. Finally, Section VI concludes the paper.

## II. Literature Review

In this Section, we provide the review of warped filters [1] and CDM technique [7].

### A. Warped Finite Impulse Response Filter

The frequency warping technique was proposed in [1] and further studied in [2-4]. Here, finite impulse response (FIR) filters are considered for the discussion. Consider an FIR filter (also called as prototype filter) with impulse response *h(n)* and its z-transform *H(z)*. Let *G(z)* be the warped version of *H(z)* obtained by replacing every delay with first order allpass structure, *A(z)* i.e.

$$G(z) = H(A(z)) \tag{1}$$

where,

$$A(z) = \left(\frac{-\alpha + z^{-1}}{1 - \alpha z^{-1}}\right) \qquad |\alpha| < 1.$$

Here, $\alpha$ is the warping coefficient which controls the warped frequency response. For $-1 < \alpha < 0$, the transformation is backward which means the transformed frequency, $f_{c\alpha}$, is smaller than the original $f_c$ and for $0 < \alpha < 1$, the effect is the reverse i.e. forward transformation which means the $f_{c\alpha}$ is higher than the original $f_c$. Various implementations are available for *A(z)* [7]. Since multiplication is the most complex operation in FPGA implementation of a digital filter, we have selected first order single-multiplier allpass structure [7] as shown in Fig. 1.

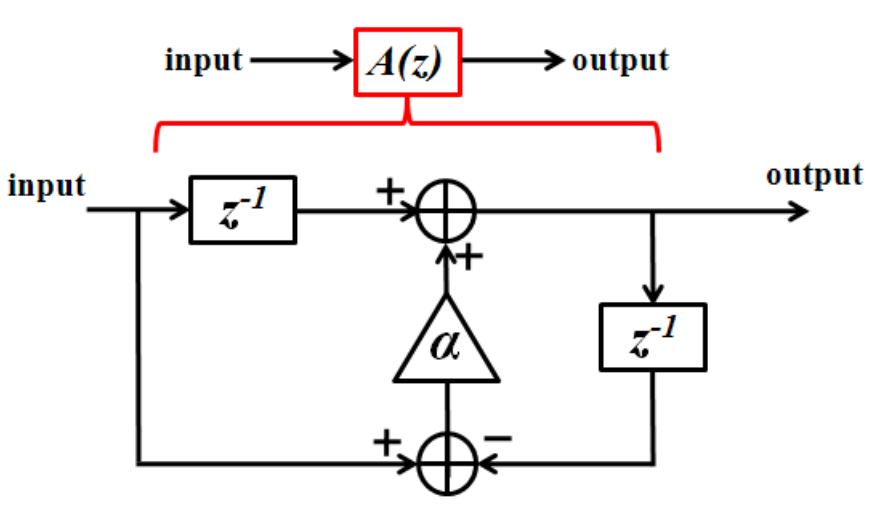

Fig. 1. First-order single-multiplier structure for *A(z)*.

The warped filter is shown in Fig. 2 [2]. It is obtained by first designing $N^{th}$ order prototype filter with coefficients $h_0$, $h_1$,.., $h_N$ in the transposed direct form and then replacing each unit delay with the *A(z)* shown in Fig. 1. The filter coefficients are fixed and warping coefficient, $\alpha$, controls the magnitude response. When $\alpha$=0, the warped filter is reduced to an FIR filter (i.e. *H(z)*) with unit delay. The warped filter has the advantages of lower filter order, improved robustness and lowered precision requirements over traditional FIR filter. The disadvantage of warped filter is increased computational complexity.

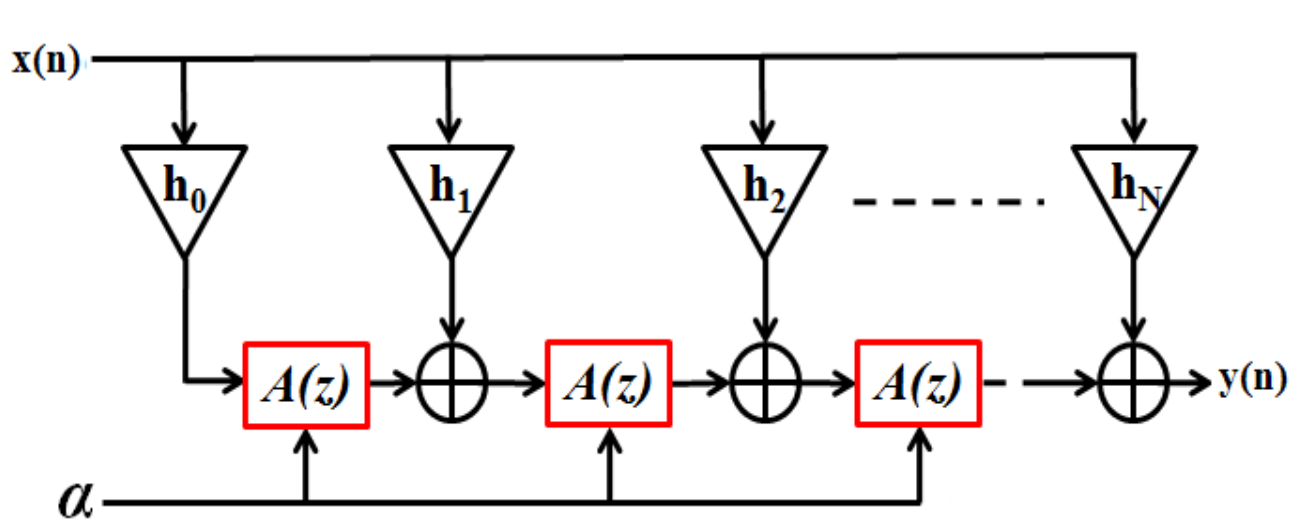

Fig. 2. Warped filter.

When the prototype filter is a bandpass filter, the output responses are variable bandpass responses. In bandpass responses, there are two variable parameters, center frequency and the bandwidth, and only one controlling parameter, $\alpha$. For example, consider the prototype bandpass filter with center frequency and bandwidth as 0.14 and 0.02 respectively. All the frequency edges mentioned here are normalized with respect to half of the sampling frequency. When $\alpha$ varies, both center frequency and bandwidth of the bandpass response varies simultaneously. The variable bandpass responses for different values of $\alpha$ are shown in Fig. 3. It can be observed that the warped filter fails to provide variable bandwidth bandpass responses for a given center frequency. The solution to this problem is achieved in this paper by using CDM technique.

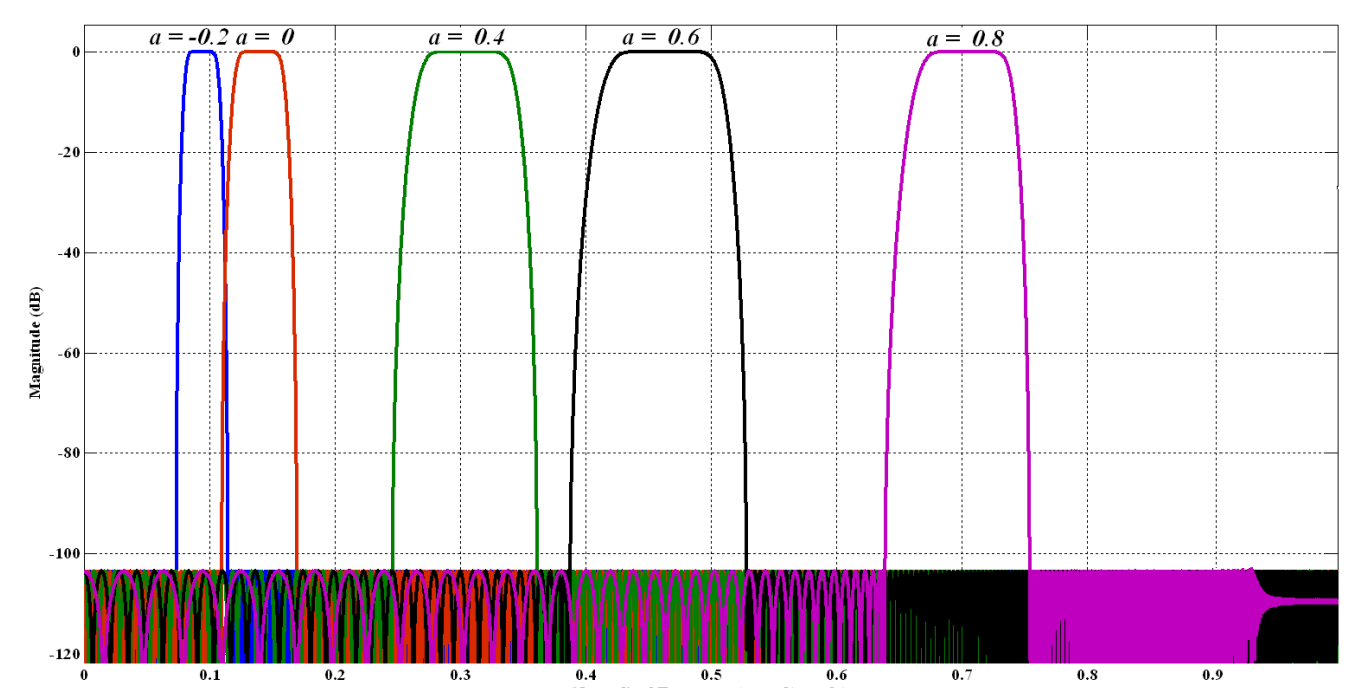

Fig. 3. Variable bandpass responses using W-FIR filter.

### B. Coefficient Decimation Technique

The CDM technique proposed in [5] provides decimated version of original frequency response whose passband width is *M* times that of prototype filter where *M* is integer decimation factor. In CDM, every $M^{th}$ coefficients of prototype filter are grouped together discarding in between coefficients. Fig. 4(b) shows the frequency responses of filter obtained using CDM from the prototype filter in fig. 4(a) for different values of $M$. It can be observed that the CDM technique alone fails to design VDFs with fine control over $f_c$, i.e. the cut-off frequencies obtained for integer incremental values of $M$ are relatively wide apart.

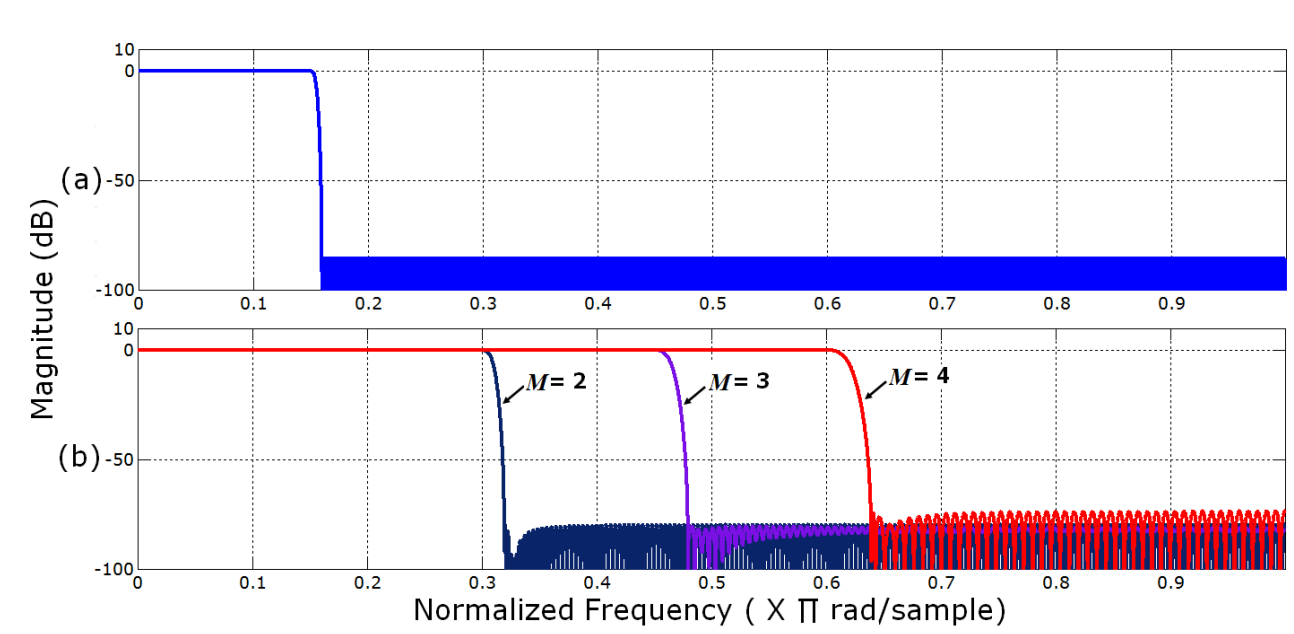

Fig. 4. (a) Frequency response of prototype filter. (b) Frequency responses of filter using CDM for different *M* from prototype filter in (a).

## III. Proposed Variable Digital Filter

The objective of the proposed VDF is to obtain variable bandwidth bandpass responses at an arbitrary center frequency using a fixed coefficient bandpass filter. In this paper, only warped FIR filter design is considered. The proposed approach can be extended to the design of warped IIR filters.

### A. Architecture of the Proposed VDF

Consider $N^{th}$ order bandpass prototype filter, $H(z)$, with center frequency, $f_{c0}$ and coefficients $h_0$, $h_1$,.., $h_N$. The filter coefficients are fixed and hence can be hard-wired. The proposed VDF architecture, obtained by combining warped filter with CDM technique, is shown in Fig. 5. The CDM is implemented using the multiplexers controlled by signal *sel_m*.

In the proposed VDF, there are two controlling parameters,

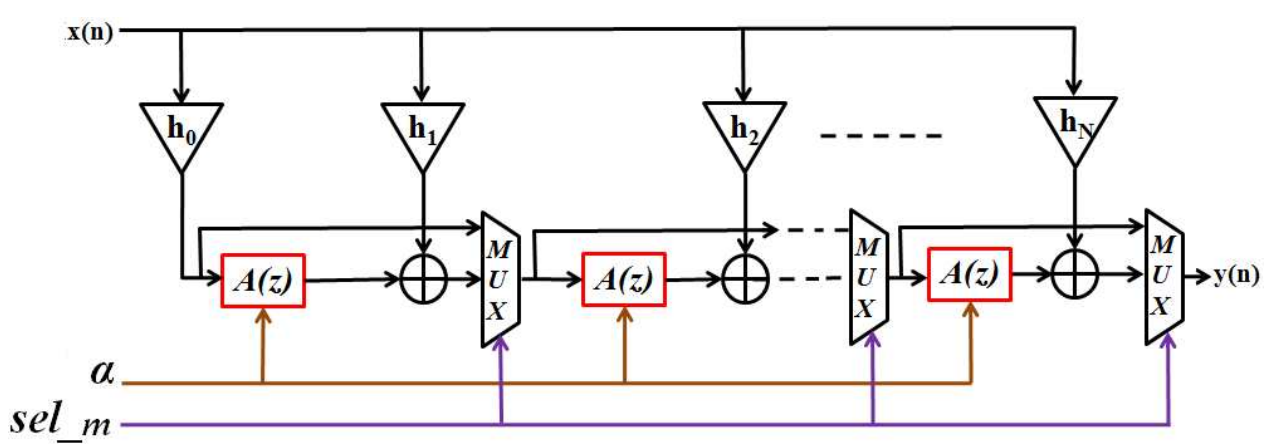

Fig. 5. Proposed VDF.

warping coefficient, $\alpha$ and decimation factor, $M$ which controls center frequency and bandwidth. The mathematical relation between desired center frequency, $f_{c\alpha}$, original center frequency, $f_{c0}$, warping coefficient $\alpha$ and decimation factor $M$ is derived below.

### B. Mathematical Derivation

The frequency response of first order causal stable real coefficient allpass filter is given by [1]

$$A(e^{j\omega}) = \frac{-\alpha + e^{-j\omega}}{1-\alpha e^{-j\omega}} \qquad (2)$$

From (2), the expression for the phase is given as [1]

$$\theta_c(\omega) = -\omega - 2\tan^{-1}\left[\frac{\alpha \sin\omega}{1-\alpha\cos\omega}\right] \qquad (3)$$

The phase delay, $\tau_p(\omega)$, of the $A(z)$ (2) is given by [1]

$$\tau_p(\omega) = \frac{-\theta_c(\omega)}{\omega} = 1 + \frac{2}{\omega}\tan^{-1}\left[\frac{\alpha\sin\omega}{1-\alpha\cos\omega}\right] \qquad (4)$$

Let the desired center frequency be $\omega_{c\alpha}$ $(= 2\pi f_{c\alpha})$. Consider prototype filter center frequency be $\omega_{c0}$ $(= 2\pi f_{c0})$. It is observed that the $\omega_{c0}$ and $\omega_{c\alpha}$ are related by

$$\omega_{c\alpha} = \frac{\omega_{c0}}{\tau_p(\omega_{c\alpha})} \times M \qquad (5)$$

$$\therefore \quad \omega_{c0} = \frac{\omega_{c\alpha}}{M} + \frac{2}{M}\tan^{-1}\left[\frac{\alpha\sin\omega_{c\alpha}}{1-\alpha\cos\omega_{c\alpha}}\right] \qquad (6)$$

where $M$ is decimation factor. Using algebraic simplification, it can be shown that

$$\alpha = \left[\frac{\tan x}{\sin\omega_{c\alpha} + \tan x \times \cos\omega_{c\alpha}}\right] \qquad (7)$$

where,

$$x = \frac{M\omega_{c1} - \omega_{c\alpha}}{2}$$

Using (7), the value of warping coefficient $\alpha$ required to obtain the desired $f_{c\alpha}$ is calculated. For the proposed VDF, $|\alpha|<1$ and $M$ takes positive integer values such as 1, 2, 3.. etc. When $M$=1, the proposed VDF is same as warped FIR filter.

The CDM operation has an inherent disadvantage of deterioration of stopband attenuation [5]. Therefore, based on the desired stopband attenuation specifications, the prototype filter should be designed with larger stopband attenuation keeping account of the deterioration of the stopband response.

## IV. Design Example

The performance of the proposed VDF is discussed in this section with the help of a suitable design example. Let the passband and stopband ripple specifications be 0.002 dB and -90 dB respectively. The center frequency, $f_c$, and bandwidth of the prototype filter are 0.14 and 0.02 respectively. The variable bandpass responses obtained using the proposed VDF for center frequencies of 0.31 and 0.71 are shown in Fig. 6. By changing $\alpha$ and $M$, center frequency and bandwidth can be varied. Thus, proposed VDF provides variable bandwidth bandpass responses for arbitrary center frequency. Similarly, variable bandstop responses can be obtained using bandstop prototype filter.

Since the proposed VDF uses the same allpass transformation as that of W-FIR filter [1] and CDM technique affects only magnitude response, phase or group delay characteristics of the proposed VDF are identical to that of warped filters [1-4]. Hence, the discussions about the phase or

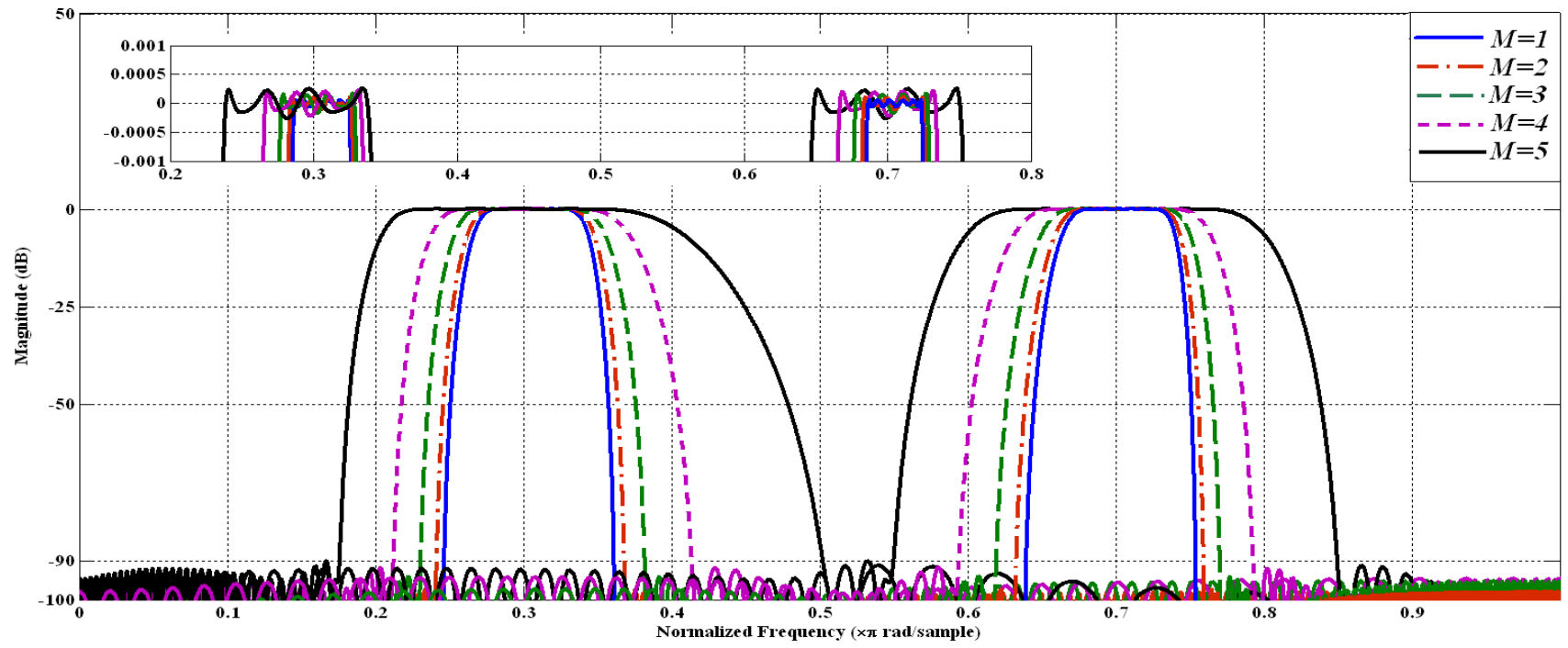
Fig. 6. Variable bandpass responses.

group delay responses are omitted here.

## V. Implementation Results

In this section, the complexity comparison in terms of total number of gate counts is done. A 16x16 bit multiplier, a 4:1 multiplexer, a word of memory and 32 bit adder were synthesized on a TSMC 0.18μm process. The Synopsys Design Compiler was used to estimate the cell area. The area in terms of gate count, as shown in Table I, was obtained by normalizing the above area values by the cell area of a two input NAND gate from the same library.

In the proposed VDF, the prototype filter needs to be overdesigned as discussed before. Thus, the order of the prototype filter for the proposed VDF is 600 compared to 550 for the warped filters in [1, 3, 4]. The warped filters in [1] require five filters in parallel to obtain variable bandpass responses identical to the proposed VDF. The warped filter in [4] is similar to [1] except filter coefficients are stored in memory instead of using parallel filters. Hence, it fails to provide online control over the bandwidth. The filter coefficients need to be updated whenever the change in bandwidth is required. The proposed VDF offers a total gate count reduction of 36% and 65% over the warped filters compared to the designs presented in [3] and [1] respectively.

## VI. Conclusion

A computationally efficient variable bandpass digital filter using warped finite impulse response filter and coefficient decimation technique is presented in this paper. The center frequency and bandwidth of this filter can be changed online without updating the filter coefficients. The design example shows that the proposed variable digital filter is simple to design and offers a total gate count reduction of 36% and 65% over the warped filters compared to the designs presented in [3] and [1] respectively.

TABLE I. Complexity Analysis

| VDFs / No. of | **Warped filter [1]** | **Warped filter with 2nd order transform [3]** | **Warped filter with memory [4]** | **Proposed VDF** |
|---|---|---|---|---|
| Multipliers | 3125 | 1375 | 825 | 901 |
| Multiplexers | 0 | 0 | 0 | 600 |
| Adders | 1750 | 3300 | 1650 | 1800 |
| Words of memory | 0 | 0 | 1275 | 0 |
| **Total gate count** | **5706250 (+65%)** | **3080000 (+36%)** | **1820750 (-6%)** | **1957700** |